\documentclass[astron]{article}
\usepackage[dvips]{epsfig}
\usepackage[colon,square,authoryear]{natbib}
\usepackage{amssymb,amsmath,amsxtra}
\usepackage{latexsym}
\usepackage[hang,small,bf]{caption}

\makeatletter
\@addtoreset{equation}{section}
\makeatother

\large\normalsize

\begin{document}

\begin{center}
{\large
\bf\Large{What local supersymmetry can do for quantum cosmology}}
\end{center}
\vspace{0.5cm}
\begin{center}
\large{Peter D.D'Eath} \\

\large{DAMTP, \\Centre for Mathematical Sciences, \\University of
Cambridge, \\ Wilberforce Road, Cambridge CB3 OWA, \\England.}
\end{center}
\vspace{1cm}
\large

\section{Introduction}\label{one}

In all approaches to quantum gravity, one makes vital use of the
classical theory, with the knowledge and intuition which this carries,
in conjunction with the quantum formulation. When, in addition, the
theory possesses local supersymmetry, this generally has profound
consequences for the nature of classical solutions, as well as for
the quantum theory.

We are used to the procedure in which a bosonic field, such as a
massless scalar field $\phi$ in flat 4-dimensional space-time, obeying
the wave equation
\begin{equation}
\square \phi=0\label{scalar}
\end{equation}
is paired with a fermionic field taken to be (say) an unprimed spinor
field $\phi^{A}$, using 2-component language \citep{penrind}. The
corresponding Weyl equation,
\begin{equation}
\nabla_{AA'}\phi^{A}=0,\label{weyl}
\end{equation}
is a system of two coupled first-order equations, which further imply
that each component of $\phi^{A}$ obeys the massless wave equation
(\ref{weyl}). These bosonic and fermionic fields may be combined (with
an auxiliary field) into a multiplet under (rigid)
supersymmetry \citep{wessbagger}. The classical fermionic field equation
(\ref{weyl}) may be viewed as a \underline{`square root'} of the original
second-order bosonic equation (\ref{scalar}).

There is a further relation here, which will be examined in the
following sections. This is with the possibility of curvature being
\underline{self-dual} in four-dimensional Riemannian
geometry \citep{atiyah}. Only in four dimensions, and only for
signature $+\,4$ (Riemannian), $0$ (ultra-hyperbolic) and $-\,4$
(equivalent to Riemannian) is this property
defined \citep{mason}. It applies both to the curvature or field
strength $F^{(a)}_{\mu\nu}$ of Yang-Mills or Maxwell theory on a
(possibly curved) Riemannian background geometry \\ \citep{penrind}, and
to the conformally invariant Weyl curvature tensor
$W_{\alpha\beta\gamma\delta}$ of the geometry, which contributes 10 of
the 20 algebraic degrees of freedom contained in the Riemann curvature tensor
$R_{\alpha\beta\gamma\delta}$. The other 10 degrees of freedom reside
in the Ricci tensor
$R_{\alpha\gamma}=g^{\beta\delta}R_{\alpha\beta\gamma\delta},$ where
$g^{\beta\delta}$ describes the inverse metric. In Einstein's theory,
the Ricci tensor corresponds to the matter source for the
curvature. The Weyl tensor $W_{\alpha\beta\gamma\delta}$ may be
thought of as describing the `vacuum' part of the gravitational field
in General Relativity.

In both the Yang-Mills and the Weyl-tensor cases, one can describe the
curvature simply in two-component spinor language \\
\citep{penrind}. The Yang-Mills field-strength tensor $F^{(a)}_{\mu\nu}=F^{(a)}_{[\mu\nu]}$
corresponds to the spinor field
\begin{equation}
F^{(a)}_{AA'BB'}=\phi^{(a)}_{AB}\varepsilon_{A'B'}+\tilde{\phi}^{(a)}_{A'B'}\varepsilon_{AB},\label{strength}
\end{equation}
where $\phi^{(a)}_{AB}=\phi^{(a)}_{(AB)}$ and
$\tilde{\phi}^{(a)}_{A'B'}=\tilde{\phi}^{(a)}_{(A'B')}$ are
symmetric. Here, $\varepsilon_{AB}$ is the unprimed alternating
spinor, and $\varepsilon_{A'B'}$ its primed counterpart. In the
present Riemannian case, the fields $\phi^{(a)}_{AB}$ and
$\tilde{\phi}^{(a)}_{A'B'}$ are independent complex quantities. If the
field strength  $F^{(a)}_{\mu\nu}$ is real, arising (say) from real
Yang-Mills potentials $v_{\mu}^{(a)}$, then each of $\phi^{(a)}_{AB}$ and
$\tilde{\phi}^{(a)}_{A'B'}$ is subject to a condition which halves the
number of real components of each. [But in the Lorentzian case, for
real $F^{(a)}_{\mu\nu}$, $\tilde{\phi}^{(a)}_{A'B'}$ will be replaced
by $\bar{\phi}^{(a)}_{A'B'}$, the complex conjugate of $\phi^{(a)}_{AB}$.]

In the Riemannian case, the Yang-Mills field strength is said to be
\underline{self-dual} if
\begin{equation}
\phi^{(a)}_{AB}=0.\label{selfdual}
\end{equation}
Similarly, an anti-self-dual field has
$\tilde{\phi}^{(a)}_{A'B'}=0$. In the case (say) of a Maxwell field in
flat Euclidean 4-space, the (anti-)self-dual condition can be written in terms
of the electric and magnetic fields as
$\underline{E}=\pm\underline{B}$. Generally, if a Yang-Mills field is
anti-self-dual, then the Yang-Mills field equations reduce to the set:
\begin{equation}
D^{B}{}_{A'}\phi^{(a)}_{AB}=0,\label{ym}
\end{equation}
where $D_{AA'}$ is the covariant derivative \citep{penrind}. This set of equations is, of course, a generalisation
of the Weyl (massless Dirac) equation (\ref{weyl}).

Regular real solutions of the anti-self-dual Yang-Mills equations
(\ref{ym}) on the four-sphere $S^{4}$ are known as
instantons \citep{eguchi}. Since (\ref{ym}) is conformally
invariant \citep{atiyah}, such a solution corresponds to a `localised'
region of Yang-Mills curvature in flat Euclidean ${\mathbb E}^{4}$, with
suitable asymptotic behaviour at large four-dimensional radius. For
the simplest non-trivial gauge group SU(2), Atiyah {\it et al.} (1978a) have, remarkably, given a construction of  the
general Yang-Mills instanton. Yang-Mills instantons can also be
described in terms of twistor theory \citep{wardwells}.

In section \ref{nobdy}, motivated by the Hartle-Hawking proposal in
quantum cosmology \citep{hawking82, hh}, we shall be led to consider Riemannian Einstein
gravity, including possibly a negative cosmological constant
$\Lambda$, in the case that the Weyl tensor is
(anti-)self-dual \citep{capovilla}. The Einstein field equations 
\begin{equation}
R_{\mu\nu}=\Lambda g_{\mu\nu}\label{fieldeq}
\end{equation}
are the conditions for an \underline{Einstein space}. The
(anti-)self-duality condition then gives a further set of equations,
closely related to the (anti-)self-dual Yang-Mills equations
(\ref{ym}) in the SU(2) case. But in the case of quantum cosmology,
the boundary conditions are usually specified on a compact connected
three-surface, such as a three-sphere $S^{3}$, in contrast to the
Yang-Mills instanton case, where they are specified at infinity. This
gravitational version is therefore more complicated. Note that a
treatment of the related boundary-value problem for Hermitian
Yang-Mills equations over complex manifolds has been given by
Donaldson (1992). In the case of 2-complex-dimensional
manifolds with K\"{a}hler metric \citep{eguchi}, this leads to anti-self-dual
Yang-Mills connections.

The corresponding purely gravitational solutions, in the case where
the boundary is at infinity (with suitable fall-off) or where the
manifold is compact without boundary, are known as gravitational
instantons \citep{hawking77, eguchi}. The (anti-)self-dual
condition on the Weyl tensor $W_{\alpha\beta\gamma\delta}$ (in the
Riemannian case) may again be described in spinor
terms \citep{penrind}: -- $W_{\alpha\beta\gamma\delta}$ corresponds to
\begin{equation}
W_{AA'BB'CC'DD'}=\Psi_{ABCD}\varepsilon_{A'B'}\varepsilon_{C'D'}+\tilde{\Psi}_{A'B'C'D'}\varepsilon_{AB}\varepsilon_{CD},\label{weyl1}
\end{equation}
where the \underline{Weyl spinors} $\Psi_{ABCD}=\Psi_{(ABCD)}$ and
$\tilde{\Psi}_{A'B'C'D'}\!=\!\tilde{\Psi}_{(A'B'C'D')}$ are again totally
symmetric. The Weyl tensor is \underline{self-dual} if 
\begin{equation}
\Psi_{ABCD}=0\label{weyl2},
\end{equation} 
and \underline{anti-self-dual} if
\begin{equation}
\tilde{\Psi}_{A'B'C'D'}=0\label{weyl2'}.
\end{equation} 
Thus, in the anti-self-dual case (say) arising in quantum cosmology,
the Ricci tensor is restricted by Eq.(\ref{fieldeq}) and the Weyl tensor by
Eq.(\ref{weyl2'}). The Bianchi identities \citep{penrind} then imply further that the remaining
Weyl spinor $\Psi_{ABCD}$ obeys
\begin{equation}
\nabla^{AA'}\Psi_{ABCD}=0.\label{weyl3}
\end{equation}   
These equations are again a generalisation of the Weyl equation
(\ref{weyl}).

Thus, at least at the formal level, there are clear resemblances
concerning supersymmetry and (anti-)self-dual classical Yang-Mills or
Einstein theory. More detail will be given in sections \ref{selfdual}-\ref{AJ}.

Turning now to the quantum theory, one has an apparent choice in
quantum cosmology between the Feynman path-integral approach \citep{hh}
and the differential approach given by Dirac's theory of the
quantisation of constrained Hamiltonian systems \cite[1950,1958a,1958b,1959,1965]{dirac}. Loosely
speaking, the latter may be thought of as a description of quantum
theories with local invariance properties, such as gauge invariance
and/or invariance under local coordinate transformations, although in
fact it is more general than that. Historically, a large amount of
work on quantum cosmology was carried out by relativists following the
pioneering work of DeWitt (1967) and Wheeler (1968)
based on the Dirac approach. The eponymous (Wheeler-DeWitt) equation
is central to the resulting quantum treatment of spatially-homogeneous
cosmologies, possibly containing bosonic matter, in which the
classical dynamics involves a (typically small) number of functions of
a time-coordinate $t$ only, and the resulting quantum field theory
reduces to a quantum-mechanical theory, with a finite number of
coordinate variables \citep{ryanshepley}. However, it has not been
possible to make sense of the second-order functional Wheeler-DeWitt
equation in the non-supersymmetric case of Einstein gravity plus
possible bosonic fields,  when the gravitational and any other bosonic
fields are allowed to have generic spatial dependence.

The path-integral approach of relevance here is that of Hartle and
Hawking (1983). There is, formally speaking, a `preferred quantum
state' for the quantum theory of (say) a spatially-compact cosmology,
where typically the coordinate variables, which become the arguments of
the wave functional, are taken to be the Riemannian three-metric
$h_{ij}$ of the compact three-manifold, together with (say) any other
bosonic fields on the three-manifold, denoted schematically by $\phi_{0}$. One then considers all possible Riemannian metrics
$g_{\mu\nu}$ and all other fields $\phi$ on all possible
four-manifolds ${\mathcal{M}}$, such that the original three-manifold
is the boundary $\partial {\mathcal{M}}$ of ${\mathcal{M}}$, and such
that the `interior' ${\mathcal{M}}$ together with its boundary
$\partial {\mathcal{M}}$, namely ${\mathcal{M}}\cup\partial
{\mathcal{M}}$ or $\bar{{\mathcal{M}}}$, is a compact
manifold-with-boundary. The three-metric and other fields inherited
from $(g_{\mu\nu},\phi)$ on the boundary must agree with the
originally prescribed $(h_{ij},\phi_{0})$. For visualisation, the
simplest example is the compact manifold $S^{3}$ (the three-sphere),
with interior the four-ball $B^{4}$. The Hartle-Hawking state
$\Psi_{HH}$, also known as the `no-boundary state', is then (formally)
described by 
\begin{equation}
\Psi_{HH}(h_{ij},\phi_{0})=\int {\mathcal{D}}g_{\mu\nu}{\mathcal{D}}\phi\exp[-I(g_{\mu\nu},\phi)/\hbar].\label{psihh}
\end{equation}
Here the functional integral is over all suitable infilling fields
$(g_{\mu\nu},\phi)$, and $I$ is the corresponding Euclidean
action \cite{hh, death}. Since the integrand is an analytic
(holomorphic) function of its arguments, this path integral may be
regarded as a giant contour integral, with the set of suitable infilling fields
deformed into the complex. The question of finding a suitable contour
for which the integral is meaningful (convergent) is a major problem
in this approach to quantum cosmology; in the above Riemannian case,
the Euclidean action $I$ is unbounded below \citep{gibbons'}, so that
the integrand in Eq.(\ref{psihh}) can become arbitrarily large and positive.

The Feynman-path-integral and Dirac-quantisation approaches are
\underline{dual} integral and differential attempts to describe the
same quantum theory, here `quantum gravity'. As shown in the book of
Feynman and Hibbs (1965), for non-relativistic quantum mechanics,
the path integral gives a wave function or quantum amplitude for a
particle to go from initial position and time
$(\underline{x}_{a},t_{a})$ to final $(\underline{x}_{b},t_{b})$, with
$t_{b}>t_{a}$, which obeys the Schr\"{o}dinger equation and also
satisfies the boundary conditions as $t_{b}\to t_{a}.$ Similarly for
the converse. Indeed Feynman's Princeton Ph.D. thesis evolved from his
continuing thought about the paper by Dirac (1933), which in
effect derived the path integral for propagation in a short
time-interval $\Delta t$. Similar dual relations even hold, somewhat
schematically, for the path-integral and differential versions of
quantum gravity \citep{hh}.

Given the above difficulties, which of these two approaches (if any)
should we use and, perhaps, trust? The bad convergence properties of
the gravitational path integral seem, at present, very difficult to
overcome. Similarly for the question of defining the second-order
Wheeler-DeWitt operator in the non-supersymmetric case. But the
Hartle-Hawking path integral provides a powerful conceptual, indeed
visual, {\it schema}. And when local supersymmetry is included,
the Wheeler-DeWitt equation is replaced by its fermionic `square
root' \citep{teitelboim}, the quantum supersymmetry
constraints \citep[1986]{death84}, which allow more sense to be made of the
quantum theory. One should expect to use both approaches together, so
far as is possible. Richard Feynman himself certainly attended to the
Dirac constrained-quantisation approach, notably in his last
substantial paper \citep{feynman}, on Yang-Mills theory in 2+1
dimensions, on which he worked for three years. Indeed, when Feynman
gave the first of the annual Dirac lectures in Cambridge, in June
1986, he remarked ``How could I refuse the invitation? -- Dirac was my
hero''. Pragmatically, anyone who has to teach a first undergraduate
course in quantum mechanics will usually base it on the
Schr\"{o}dinger equation (the differential approach). But, of course,
there is nothing to stop them from inducting the students {\it via} the
path-integral approach. Maybe this has been tried, at least at
CalTech!

\section{No-Boundary State}\label{nobdy}

To expand on the description (\ref{psihh}) of the Hartle-Hawking state, we
will need later to be able to include fermions with the gravitational
and other bosonic fields, in the Riemannian context. This requires the
introduction of an orthonormal tetrad $e^{a}{}_{\mu}$ of one-forms;
here $a=0,1,2,3$ labels the four orthonormal co-vectors
$e^{a}{}_{\mu}$, while $\mu$ is the `space-time' or `world' index. One
has, by orthogonality and completeness:
\begin{equation}
g^{\mu\nu}e^{a}{}_{\mu}e^{b}{}_{\nu}=\delta^{ab},\,\,\,\,\delta_{ab}e^{a}{}_{\mu}e^{b}{}_{\nu}=g_{\mu\nu}.\label{ortho}
\end{equation}
Thus $e^{a}{}_{\mu}$ is a `square root' of the metric $g_{\mu\nu}$,
non-unique up to local SO(4) rotations acting on the internal index
$a$.

In this case, instead of specifying the intrinsic three-metric
$h_{ij}$ on the boundary $(i,j,...=1,2,3)$, one would specify the four
spatial one-forms $e^{a}{}_{i}$, with
\begin{equation}
h_{ij}=\delta_{ab}e^{a}{}_{i}e^{b}{}_{j}.\label{hij}
\end{equation}
This seems a redundant description, since only three co-vectors
$e^{a}{}_{i}$ are needed to take the `square root' of $h_{ij}$ in
Eq.(\ref{hij}). On and near the boundary surface, it is valid to work in
the `time gauge' \citep{nelson}, using a \underline{triad}
$e^{a}{}_{i}$ $(a=1,2,3)$ to obey Eq.(\ref{hij}). Equivalently, one
imposes $e^{0}{}_{i}=0$ as a gauge condition. Instead of integrating
over all Riemannian four-metrics $g_{\mu\nu}$ in Eq.(\ref{psihh}), one
integrates over all $e^{a}{}_{\mu}$, each of which corresponds to a
Riemannian metric by Eq.(\ref{ortho}).

The gravitational part of the Euclidean action $I$ in Eq.(\ref{psihh})
is \citep{gibbhaw77}
\begin{equation}
I_{grav}=-\frac{1}{2\kappa^{2}}\int_{{\mathcal{M}}}d^{4}x\,g^{\frac{1}{2}}(R-2\Lambda)-\frac{1}{\kappa^{2}}\int_{\partial{\mathcal{M}}}d^{3}x\,h^{\frac{1}{2}}tr
K\label{trk}.
\end{equation} 
Here $\kappa^{2}=8\pi$, $g=det(g_{\mu\nu})=[det(e^{a}{}_{\mu})]^{2}$,
$R$ is the four-dimensional Ricci scalar, $\Lambda$ is the
cosmological constant and $tr K=h^{ij}K_{ij}$, where $K_{ij}$ is the
second fundamental form (or extrinsic curvature) of the
boundary \citep{mtw, death}.

In the path integral (\ref{psihh}), one expects to sum over all
four-manifolds ${\mathcal{M}}$, of different topology, which have the
prescribed three-manifold as boundary $\partial {\mathcal{M}}$; for
each topologically different ${\mathcal{M}}$, one then integrates over
metrics $g_{\mu\nu}$ or tetrads $e^{a}{}_{\mu}$. For each choice of
${\mathcal{M}}$, one can ask whether there are any solutions of the
classical field equations, for the given boundary data $h_{ij}$ or
$e^{a}{}_{i}$, other bosonic data $\phi_{0}$ and possible fermionic
data. In the simplest case of gravity without matter (for
definiteness), there may be zero, one, two, ... real Riemannian
solutions of Eq.(\ref{fieldeq}) for a given topology ${\mathcal{M}}$. Of
course, for the same ${\mathcal{M}}$ and real boundary data $h_{ij}$,
there may be a larger number of \underline{complex} classical
solutions $g_{\mu\nu}$.

Suppose, again for definiteness, that we again have gravity without
matter, and that there is a unique classical solution $g_{\mu\nu}$ (up
to coordinate transformation) which is Riemannian (and hence real) on
a particular four-manifold ${\mathcal{M}}_{0},$ corresponding to the
boundary data $h_{ij}(\underline{x})$. Further, suppose that there are
no other classical solutions on any other manifold
${\mathcal{M}}$. Then, were the path integral (\ref{psihh}) to be
meaningful, one would expect to have a \underline{semi-classical
expansion} of the Hartle-Hawking state, of the form
\begin{equation}
\Psi_{HH}[h_{ij}(\underline{x})]\sim (A_{0}+\hbar A_{1}+\hbar^{2}A_{2}+...)\exp(-I_{class}/\hbar).\label{asymp}
\end{equation} 
Here the wave function $\Psi_{HH}$, the `one-loop factor' $A_{0}$,
`two-loop factor' $A_{1}$, ... and the Euclidean action $I_{class}$ of
the classical solution [as in Eq.(\ref{trk})] are all functionals of
$h_{ij}(\underline{x})$. Technically, one might expect that such an
expansion, if it existed, would only be an asymptotic expansion valid
in the limit as $\hbar\to 0_{+}.$ Even in non-relativistic quantum
mechanics, semi-classical expansions are typically only asymptotic but
not convergent \citep{itzykson}. Note also that, if well-posed
fermionic boundary data are included, and there is a unique
corresponding coupled bosonic-fermionic classical solution, then one
expects again a semi-classical wave function $\Psi_{HH}$ of the
boundary data, of the form (\ref{asymp}), except that each of
$I_{class}$, $A_{0}$, $A_{1}$, $A_{2}$,... will be a functional of the
complete bosonic and fermionic boundary data.

\section{The classical Riemannian boundary-value \\ problem}\label{riem}

\subsection{The general boundary problem}\label{riem1}

As follows from section \ref{nobdy}, it is important to understand the
nature of the Riemannian boundary-value problem for Einstein gravity,
possibly including a $\Lambda$-term, matter fields and local
supersymmetry. Only very partial results are available in the
generic case for which the boundary data has no
symmetries. Reula (1987) proved an existence theorem for the
vacuum Riemannian Einstein equations $(\Lambda = 0)$ on a slab-like
region, where suitable data on two parallel planes enclosing a slab of
Euclidean ${\mathbb E}^{4}$ are slightly perturbed. For weak perturbations of a
suitable known compact manifold-with-boundary, the case $\Lambda\leq
0$ was studied by Schlenker (1998). To fix one's intuition,
consider the case in which the unperturbed boundary is a metric
three-sphere $S^{3}$, bounding part of flat ${\mathbb E}^{4}$ (if
$\Lambda = 0$) or of a hyperbolic space ${\mathbb H}^{4}$ (if $\Lambda < 0$). Then any
sufficiently weak perturbation of the boundary metric $h_{ij}$ yields
a corresponding (perturbed but non-linear) interior solution
$g_{\mu\nu}$ for the 4-metric, obeying $R_{\mu\nu}=\Lambda
g_{\mu\nu}.$ 

Boundary-value problems `at infinity' have also been studied, for
$\Lambda <0$, when a 4-dimensional Riemannian geometry can be given a
conformal infinity \citep{grahamlee}. The canonical example is
hyperbolic space ${\mathbb H}^{4}$, with its metric of constant curvature, here
normalised such that $\Lambda=-12$. Let $ds_{b}^{2}$ denote the flat
Euclidean metric on ${\mathbb R}^{4}$, expressed in terms of Cartesian
coordinates $x^{\mu}\,(\mu=0,1,2,3)$; then define the conformal
function $\rho=\frac{1}{2}(1-|x|^{2})$ on the unit ball $B^{4}\subset{\mathbb R}^{4}$. The \underline{conformal metric} $ds^{2}=\rho^{-2}ds_{b}^{2}$
is the hyperbolic metric on the open set $B^{4}$, and the `conformal
metric at infinity $(|x|=1)$' can be taken to be the standard metric
$H_{ij}$ on $S^{3}$. Graham and Lee (1991) have shown that
this `conformal Einstein' problem is also well-behaved for small
perturbations of the unit-sphere metric $H_{ij}$ on the
\underline{conformal boundary} $S^{3}$. As in the previous paragraph,
for 3-metrics $h_{ij}$ sufficiently close to $H_{ij}$, there is a
corresponding conformal metric on the interior $B^{4}$, close to the
unperturbed hyperbolic metric.

The case with conformal infinity, imposing the Einstein condition
$R_{\mu\nu}=\Lambda g_{\mu\nu}$ with $\Lambda <0$, has also been
studied subject to the additional requirement of (say) self-duality of
the Weyl tensor, $\Psi_{ABCD}=0$ [Eq.(\ref{weyl2'})]. LeBrun (1982) has shown that, when the conformal infinity $\partial {\mathcal{M}}$
is a real-analytic 3-manifold with conformal metric $h_{ij}$, then, in
a neighbourhood of $\partial {\mathcal{M}}$, there is a conformal
4-metric $g_{\mu\nu}$ on a real-analytic 4-manifold ${\mathcal{M}}$,
satisfying the Einstein equations with $\Lambda <0$ and self-dual Weyl
curvature. The real-analytic condition on the conformal boundary
$\partial {\mathcal{M}}$ is essential, since the Einstein-space
condition $R_{\mu\nu}=\Lambda g_{\mu\nu}$ together with self-duality
imply that the 4-manifold must be real-analytic \citep{atiyah}. Further, LeBrun (1991) has shown
that there is an infinite-dimensional space of conformal metrics
$h_{ij}$ on $S^{3}$ which bound \underline{complete} Einstein metrics
on the 4-ball, with (anti-)self-dual Weyl curvature; that is, $S^{3}$
is again conformal infinity, but now the result is not just local, in a
neighbourhood of the $S^{3}$ boundary, but extends smoothly across the
interior, the 4-ball.

Finally, note that LeBrun (1982) also proved a local result in
which the conformal infinity $\partial {\mathcal{M}}$ is taken to be a
suitable complex 3-manifold with given holomorphic
metric \citep{wells}, and a complex 4-manifold ${\mathcal{M}}$ in a
neighbourhood of $\partial {\mathcal{M}}$ is then guaranteed to exist,
with holomorphic metric obeying $R_{\mu\nu}=\Lambda g_{\mu\nu}$ and
$\Lambda\neq 0$ (possibly complex), together with self-duality of the
Weyl tensor. This is of potential interest in quantum gravity, since,
as in section \ref{one}, the Hartle-Hawking path integral is a contour
integral, and there may be stationary points (classical solutions)
with holomorphic 4-metrics; further, one would expect to be able to
continue the boundary data, such as $h_{ij}$, into the complex (i.e.,
holomorphically).

\subsection{Example - biaxial Riemannian Bianchi-IX models}\label{example}

As an example, consider the family of Riemannian 4-metrics with
isometry group SU(2)$\times $U(1), given (locally in the coordinate
$r$) by:
\begin{equation}
ds^{2}=dr^{2}+a^{2}(r)(\sigma_{1}^{2}+\sigma_{2}^{2})+b^{2}(r)\sigma_{3}^{2}\label{ds2}.
\end{equation}
Here, $a(r)$ and $b(r)$ are two functions of the `radial' coordinate
$r$, and $\{\sigma_{1}, \sigma_{2}, \sigma_{3}\}$ denotes the basis of
left-invariant 1-forms (co-vector fields) on the three-sphere $S^{3}$,
regarded as the group SU(2), with the conventions
of \citep{eguchi}. The more general triaxial Bianchi-IX
metric \citep{kramer} -- see below -- would have three different
functions multiplying $\sigma_{1}^{2}$, $\sigma_{2}^{2}$ and
$\sigma_{3}^{2}$ in Eq.(\ref{ds2}).

In the biaxial case, the boundary data at a value $r=r_{0}$ are taken
to be the intrinsic 3-metric 
\begin{equation}
ds^{2}=a^{2}(r_{0})(\sigma_{1}^{2}+\sigma_{2}^{2})+b^{2}(r_{0})\sigma_{3}^{2}\label{ds2'},
\end{equation}
determined by the positive numbers $a^{2}(r_{0})$ and
$b^{2}(r_{0})$. This gives a `squashed 3-sphere' or \underline{Berger
sphere}. Thus one wishes to find a regular solution of the
Einstein-$\Lambda$ field equations (\ref{fieldeq}) in the interior
${\mathcal{M}}$ of the boundary $\partial {\mathcal{M}}\cong S^{3}$
(denoting `$\partial {\mathcal{M}}$ is diffeomorphic to $S^{3}$'),
subject to the boundary data (\ref{ds2'}). There are two possible ways in
which such a 4-geometry could close in a regular way as $r$ is
decreased from $r_{0}$, to give a compact manifold-with-boundary
${\mathcal{M}}\cup \partial {\mathcal{M}}=\bar{{\mathcal{M}}}$. Either
${\mathcal{M}}$ is (diffeomorphically) a 4-ball $B^{4}$, with standard
polar-coordinate behaviour
\begin{equation}
a(r)\sim r,\,\,\,\, b(r)\sim r\,\,\,\mathrm{as}\,\, r\to 0\label{limit}
\end{equation}
near the `centre' $r=0$ of the 4-ball. Or ${\mathcal{M}}$ has a more
complicated topology, still with boundary $\partial
{\mathcal{M}}\cong S^{3}$, such that
\begin{equation}
a(r)\to c (\mathrm{constant}>0),\,\,\,\, b(r)\sim r\,\,\,\mathrm{as}\,\, r\to 0\label{limit'}.
\end{equation}
Here the 4-metric degenerates to the metric of a round 2-sphere
$S^{2}$, as $r\to 0$. The first case is described as \underline{NUT
behaviour} as $r\to 0$, and the second as \underline{BOLT
behaviour} \citep{gibbhaw79, eguchi}. In both cases,
the apparent singularity at $r=0$ is a removable coordinate singularity.

The general Riemannian solution of the Einstein field equations
$R_{\mu\nu}=\Lambda g_{\mu\nu}$ for biaxial Bianchi-IX metrics can be
written in the form \citep{gibbhaw79, gibbpope}
\begin{equation}
ds^{2}=\frac{(\rho^{2}-L^{2})}{4\Delta}d\rho^{2}+(\rho^{2}-L^{2})(\sigma_{1}^{2}+\sigma_{2}^{2})+\frac{4L^{2}\Delta}{(\rho^{2}-L^{2})}\sigma_{3}^{2},\label{ds2''}
\end{equation} 
where
\begin{equation}
\Delta=\rho^{2}-2M\rho+L^{2}+\frac{1}{4}\Lambda(L^{4}+2L^{2}\rho^{2}-\frac{1}{3}\rho^{4}).\label{delta}
\end{equation}
This 2-parameter family of metrics, labelled (for given $\Lambda$) by
the constants $L, M,$ is known as the Taub-NUT-(anti)de Sitter family.

It was found by Jensen {\it et al.} (1991) that a 4-geometry in
this family has NUT behaviour (near $\rho^{2}=L^{2}$) precisely when
one of the relations
\begin{equation}
M=\pm L(1+\frac{4}{3}\Lambda L^{2})\label{M}
\end{equation} 
holds. Further \citep*{gibbpope}, these are the conditions for (anti-)self-duality of the Weyl tensor. In the classical NUT boundary-value
problem, positive values of $A=a^{2}(r_{0}), B=b^{2}(r_{0})$ are
specified on the boundary $\partial {\mathcal{M}}$, and the geometry
must fill in smoothly on a 4-ball interior, subject to the Einstein
equations $R_{\mu\nu}=\Lambda g_{\mu\nu}$. As remarked by Jensen {\it et
al.} (1991), NUT regularity corresponds to one further
requirement, given by a cubic equation, beyond Eq.(\ref{M}). This is
investigated further in \\ \citep*{akbar}; see also Chamblin {\it et
al.} (1999). Taking (say) the anti-self-dual case in
Eq.(\ref{M}), assuming also $\Lambda <0$ for the sake of argument, and
given positive boundary values $(A,B)$, the cubic leads to three
regular NUT solutions (counting multiplicity). Depending on the values
$(A,B)$, from zero to three of these are \underline{real} Riemannian
solutions of the type (\ref{ds2''}). The remaining NUT solutions are inevitably
complex (holomorphic) geometries. In the physically interesting limit,
where both `cosmological' (radii)$^{2}$ $A$ and $B$ are large and
comparable, all three solutions are real. In the Hartle-Hawking path
integral (\ref{psihh}) and its semi-classical expansion (\ref{asymp}), with
Euclidean action $I_{grav}(\,)$, this would give an estimate, say for the
isotropic case $A=B$:
\begin{equation}
I_{class}\sim-\frac{\pi}{12 |\Lambda |}A^{2}\,\,\,\mathrm{as}\,\,A\to\infty,\label{iclass}
\end{equation} 
with 
\begin{equation}
\Psi_{HH}\sim (\mathrm{slowly\, varying\, prefactor})\times\exp(-I_{class}/\hbar)\label{psiasym}.
\end{equation}
In such an Einstein-negative-$\Lambda$ model, without further matter,
the relative probability of finding a universe with a given $A$ would
increase enormously with $A$.

If instead $\Lambda =0$, the solution (\ref{ds2''},6) reduces to the
Euclidean Taub-NUT solution \citep{hawking77}. For $\Lambda >0$, one may
visualise the isotropic case $A=B$, with a metric 4-sphere $S^{4}$ as
Riemannian solution, the radius being determined in terms of
$\Lambda$. When the (radii)$^{2}$ $A$ and $B$ become too large
relative to $\Lambda^{-1}$, there will be no real Riemannian
solution \citep{jensen}, but only complex (holomorphic) geometries.

The alternative regular BOLT solutions are studied in \\ \citep{akbar'},
particularly in the case $\Lambda <0$, for given positive boundary
data $(A,B)$. These solutions \underline{do not} have an (anti-)self-dual
Weyl tensor. Further, their topology is more complicated than that of
the 4-ball $B^{4}$ -- the simplest way of filling in an $S^{3}$. This
difference can be seen, for example, by computing the topological
invariants $\chi$, the Euler invariant, and $\tau$, the Pontryagin
number, each of which is given by a volume integral quadratic in the
Riemann tensor, together with a suitable surface
integral \citep{eguchi}. For the 4-ball, one has $\chi =1, \tau =0$;
for a BOLT solution, $\chi =2, \tau =-1$. The problem of finding BOLT
solutions depends on studying a seventh-degree polynomial! The number
of \underline{real} regular BOLT solutions, for given positive
boundary data $(A,B)$, must be twice a strictly positive odd number;
other solutions are necessarily complex. When the boundary is not too
anisotropic, i.e., when $A$ and $B$ are sufficiently close to one
another, there are exactly two regular BOLT solutions.

Of course, one could in principle study the corresponding much more
elaborate triaxial boundary-value problem. This has at least been done
for the case of a conformal boundary at infinity, as in section
(\ref{riem1}), with conformal 3-metric of triaxial Bianchi-IX
type \citep{hitchin}. The solutions involve Painlev\'{e}'s sixth
equation \\ \citep{mason}; see also \citep{tod}.

\section{Self-duality}\label{selfdual}

\subsection{Hamiltonian approach; Ashtekar variables}\label{ashvar}

Consider now a Hamiltonian treatment of Einstein gravity with a
$\Lambda$-term, modified for use in the Riemannian or `imaginary-time'
case. Since we shall later include fermions, a tetrad (or triad)
description of the geometry must be used, as in Eqs.(\ref{ortho},2),
except that we shall use spinor-valued one-forms $e^{AA'}{}_{\mu}$
instead of the tetrad $e^{a}{}_{\mu}$. Here 
\begin{equation}
e^{AA'}{}_{\mu}=\sigma_{a}^{AA'}e^{a}{}_{\mu},
\end{equation}
where $\sigma_{a}^{AA'}$ are appropriate Infeld-van der Waerden
translation symbols \citep{penrind}. The spatial 3-metric $h_{ij}$
is given by
\begin{equation}
h_{ij}=-e_{AA'i}e^{AA'}{}_{j},
\end{equation}
where the spinor-valued spatial one-forms $e^{AA'}{}_{i}$ are regarded
as the coordinate variables in a `traditional' Hamiltonian
treatment. The \underline{Lorentzian normal vector} $n^{\mu}$ has
spinor version $n^{AA'}$, which is determined once the $e^{AA'}{}_{i}$
are known \citep{death}. In our Riemannian context, the corresponding
\underline{Euclidean normal vector} $_{e}n^{\mu}$ corresponds to 
\begin{equation}
_{e}n^{AA'}=-in^{AA'}.
\end{equation}

In the `time gauge' of section \ref{nobdy}, one has only a triad
$e^{a}{}_{i}(a=1,2,3)$, and the remaining one-form $e^{0}{}_{\mu}$ is
constrained by $e^{0}{}_{i}=0$; equivalently
$_{e}n^{\mu}=\delta^{a}_{0}.$ The local invariance group of the theory
becomes SO(3) in the triad version.

For Riemannian 4-geometries, the torsion-free connection is given by
the \underline{connection 1-forms} $\omega^{ab}{}_{\mu}=\omega^{[ab]}{}_{\mu}$ \citep{death}. In spinor language, these correspond to 
\begin{equation}
\omega^{AA'BB'}{}_{\mu}=\omega^{AB}{}_{\mu}\varepsilon^{A'B'}+\tilde{\omega}^{A'B'}{}_{\mu}\varepsilon^{AB},\label{omega}
\end{equation}
where $\omega^{AB}{}_{\mu}=\omega^{(AB)}{}_{\mu}$ is a set of 1-forms
taking values in the Lie algebra $\mathfrak s\mathfrak u$(2) of the group
SU(2); similarly for the independent quantity
$\tilde{\omega}^{A'B'}{}_{\mu}=\tilde{\omega}^{(A'B')}{}_{\mu}$, with
a different copy of SU(2). Then the curvature is described by the
2-forms $R^{AB}{}_{\mu\nu}=R^{(AB)}{}_{[\mu\nu]}$, defined by
\begin{equation}
R^{AB}{}_{\mu\nu}=2(\partial_{[\mu}\omega^{AB}{}_{\nu]}+\omega^{A}{}_{C[\mu}\omega^{CB}{}_{\nu]}),\label{rab}
\end{equation}
and a corresponding $\tilde{R}^{A'B'}{}_{\mu\nu}.$ In the language of
forms \citep{eguchi}, the spinor-valued 2-form $R^{AB}$ is defined
equivalently as 
\begin{equation}
R^{AB}=d\omega^{AB}+\omega_{A}{}^{C}\wedge\omega_{C}{}^{B}\label{rab1}
\end{equation}
and corresponds to the anti-self-dual part of the Riemann
tensor. Similarly, $\tilde{R}^{A'B'}$ corresponds to the self-dual
part.

In the Hamiltonian formulation of Ashtekar (1986, 1987, 1988, \\ 1991), one
defines the spatial spinor-valued 1-forms $\sigma^{AB}{}_{i}=\sigma^{(AB)}{}_{i}$ as
\begin{equation}
\sigma^{AB}{}_{i}=\sqrt{2}ie^{A}{}_{A'i}n^{BA'}.
\end{equation}
These can equally be described, in the time gauge, in terms of the
spatial triad $e^{a}{}_{i}$, the translation symbols
$\sigma^{AA'}_{a}$ and the unit matrix $\delta^{AA'}.$ Then, with
$\sigma^{ABi}=h^{ij}\sigma^{AB}{}_{j}$, one defines the density
\begin{equation}
\tilde{\sigma}^{ABi}=h^{\frac{1}{2}}\sigma^{ABi},\label{density}
\end{equation}
where $h=det(h_{ij})$. The \underline{Ashtekar canonical variables}
are then $\tilde{\sigma}^{ABi}$ and $\omega_{ABi}$, the spatial part
of the unprimed connection 1-forms with spinor indices lowered. It can
be verified that these are canonically conjugate. Of course, since,
they contain only unprimed spinor indices, they are very well adapted
for a description of (anti-)self-duality. For this purpose, we shall
also need the spinor-valued 2-form
\begin{equation}
\Sigma_{AB}=e_{A}{}^{A'}\wedge e_{BA'},\label{Sigma}
\end{equation}
obeying $\Sigma_{AB}=\Sigma_{(AB)}$.

In the Hamiltonian approach \\ \citep{ash, jacob, capovilla},
the action can be decomposed in terms of the spatial `coordinate
variables' $\omega_{ABi}=\omega_{(AB)i}$ and `momentum variables'
$\tilde{\sigma}^{ABi}=\tilde{\sigma}^{(AB)i}$, together with the
Lagrange multipliers \underline{N} (lapse), $N^{i}$ (shift) and
$\omega_{AB0}$ [specifying local SU(2) transformations]. The
(Lorentzian) action $S$ has the form 
\begin{equation}
S=\int Tr[\tilde{\sigma}^{i}\dot{\omega}_{i}+(\underline{N}\tilde{\sigma}^{i}\tilde{\sigma}^{j}(R_{ij}-\frac{1}{3}\Lambda\Sigma_{ij})+N^{i}\tilde{\sigma}^{j}R{ij}+\omega_{0}D_{i}\tilde{\sigma}^{i})].\label{S}
\end{equation} 
Here, all spinor indices have been suppressed, but spatial indices are
left explicit. The conventions $(MN)_{A}{}^{C}=M_{A}{}^{B}N_{B}{}^{C}$
and $Tr(M_{A}{}^{B})=M_{A}{}^{A}$ are being used. The spatial
curvature 2-forms $R^{AB}{}_{ij}=R^{(AB)}{}_{[ij]}$ are constructed as
in Eq.(\ref{rab}) from the spatial connection 1-forms
$\omega^{AB}{}_{i}$, and $D_{i}$ denotes the spatial covariant
derivative. From variation of the Lagrange multipliers, one finds as
usual that each of their coefficients vanish, giving the
\underline{constraint equations} which restrict the form of the allowed
data $(\omega_{ABi},\tilde{\sigma}^{ABi})$ for classical solutions
(whether anti-self-dual or more general). Further, the spatial 2-forms
$\Sigma^{(AB)}{}_{[ij]}$ of Eq.(\ref{Sigma}) are related to the variables
$\tilde{\sigma}^{ABi}$ by
\begin{equation}
\tilde{\sigma}^{ABi}=\varepsilon^{ijk}\Sigma^{AB}{}_{jk}.
\end{equation} 

The generalisation of Ashtekar's approach to supergravity was given by
Jacobson \nocite{jacob}(1988), and will be used in section \ref{AJ}.

\subsection{Non-zero $\Lambda$: the anti-self-dual case and the
Chern-Simons functional}\label{chern}

\indent

It was shown in \citep[1991; Samuel 1988]{capovilla} that the anti-self-dual Einstein field equations (\ref{fieldeq},9), with a
\underline{non-zero} \underline{cosmological constant} $\Lambda$, can be re-expressed in terms of the
(4-dimension\-al) 2-form $\Sigma^{AB}= \Sigma^{(AB)}$. Note first that,
for any set of orthonormal 1-forms $e^{AA'}{}_{\mu}$, the 2-forms
$\Sigma^{AB}$ defined in Eq.(\ref{Sigma}) automatically obey
\begin{equation}
\Sigma^{(AB}\wedge\Sigma^{CD)}=0.\label{prodsigma}
\end{equation} 
Equivalently, for a \underline{real} SO(3) triad $e^{a}{}_{\mu}$,
where $a=1,2,3$ here, the conditions (\ref{prodsigma}) read
\begin{equation}
\Sigma^{a}\wedge\Sigma^{b}-\frac{1}{3}\delta^{ab}\Sigma^{c}\wedge\Sigma_{c}=0.\label{sigmawedge'}
\end{equation} 
In the case of anti-self-dual Weyl curvature ($\Psi_{ABCD}=0)$, the
Einstein field equations reduce to
\begin{equation}
R^{AB}=\frac{1}{3}\Lambda \Sigma^{AB}.
\end{equation}
 
Conversely, these authors show that, given any
$\mathfrak s\mathfrak u$(2)-valued unprimed connection 1-form $\omega^{AB}=\omega^{(AB)}$, with corresponding curvature 2-forms
$R^{AB}=R^{(AB)}$ defined by Eq.(\ref{rab1}), it is sufficient (locally)
that the $R^{AB}$ further obey the \underline{algebraic conditions}
\begin{equation}
R^{(AB}\wedge R^{CD)}=0,\label{rabwedge}
\end{equation} 
or the equivalent of Eq.(\ref{sigmawedge'}), with $R^{a}$ replacing $\Sigma^{a}$,
in the real SO(3) triad case. Then, defining $\Sigma^{AB}$ by the
inverse of Eq.(\ref{rabwedge}):
\begin{equation}
\Sigma^{AB}=3\Lambda^{-1}R^{AB,}
\end{equation} 
it is shown that (locally) this defines, {\it via} Eq.(\ref{Sigma}),
a metric which obeys the Einstein-$\Lambda$ equations, with
anti-self-dual Weyl curvature.

The Hamiltonian approach with $\Lambda\neq 0$, taking canonical
variables $(\omega_{ABi},\tilde{\sigma}^{ABi})$ with action $S$ given
by Eq.(\ref{S}), has been further discussed by \citep{koshti}. A
necessary and sufficient condition for an initial-data set to
correspond locally to a solution of the Einstein equations with
anti-self-dual Weyl curvature is that
\begin{equation}
\tilde{\sigma}^{ABi}=\frac{-3}{\Lambda}\tilde{B}^{ABi}.\label{tildesigma}
\end{equation} 
Here, $\tilde{B}^{ABi}$ is a densitised version of the magnetic part
of the Weyl tensor \citep{mtw}, defined by
\begin{equation}
\tilde{B}^{ABi}=\frac{1}{2}\varepsilon^{ijk}R^{AB}{}_{jk},
\end{equation} 
where $\varepsilon^{ijk}$ is the alternating symbol in 3 dimensions,
and $R^{AB}{}_{jk}=R^{(AB)}{}_{[jk]}$ gives, as usual, the spatial part
of the curvature 2-form, following Eq.(\ref{rab}).

The evolution equations are most easily described in terms of the
equivalent variables $(\omega_{ai},\tilde{\sigma}^{ai})$, where
$a=1,2,3$ is a local SO(3) index. Recall that $\omega_{ai}$ and
$\tilde{\sigma}^{ai}$ are defined by
$\omega_{ai}=\Sigma_{a}^{AB}\omega_{ABi},$
$\tilde{\sigma}^{ai}=\Sigma^{a}_{AB}\omega^{ABi}$, where
$\Sigma_{a}^{AB}$ and $\Sigma^{a}_{AB}$ are the triad Infeld-van der
Waerden translation symbols. Then, in the anti-self-dual Riemannian
case, the normal derivative of $\omega_{ai}$ is given by
\begin{equation}  
\dot{\omega}_{ai}=\left(\frac{3}{4\Lambda}\right)\underline{N}\varepsilon_{ijk}\varepsilon_{abc}\tilde{B}^{bj}\tilde{B}^{ck}.\label{evolution}
\end{equation} 
Here, in the language of Eq.(\ref{S}), only the `lapse' Lagrange
multiplier \underline{N} is taken non-zero. Hence, if $\omega_{ai}$ is
specified on a hypersurface, then, assuming anti-self-duality, the
conjugate variable $\tilde{\sigma}^{ai}$ is determined by
Eq.(\ref{tildesigma}). The evolution of $\omega_{ai}$ away from the
hypersurface is then determined by solving the set of partial
differential equations (\ref{evolution}), which involves no more than first derivatives of
$\omega_{ai}$, in the form $\dot{\omega}_{ai}$ and
$\varepsilon^{ijk}\partial\omega_{aj}/\partial x^{k}$, the latter
quadratically. Away from the bounding hypersurface, the conjugate
variables $\tilde{\sigma}^{ai}$ continue to be given in terms of
$\omega_{ai}$ by Eq.(\ref{tildesigma}).

As usual for Hamiltonian systems with first-order evolution for the
`coordinates' alone (here $\omega_{ai}$), the classical action
$I[\omega_{ai}]$, regarded as a functional of the boundary data
$\omega_{ai},$ is the principal generating
function \citep{arnold, goldstein, landau}, with (in spinor language):
\begin{equation}  
\frac{\delta I}{\delta \omega_{ABi}}=\tilde{\sigma}^{ABi},
\end{equation} 
together with the correct evolution equations. Up to an additive
constant, the classical action $I[\omega_{ai}]$ is precisely the
\underline{Chern-Simons action}
\begin{equation}  
I_{CS}[\omega_{ai}]=\frac{-3}{2\Lambda}\int\varepsilon^{ijk}[\omega^{AB}{}_{i}(\partial_{j}\omega_{ABk})+\frac{2}{3}\omega^{AB}{}_{i}\omega_{B}{}^{C}{}_{j}\omega_{CAk}],\label{ics'}
\end{equation} 
as studied in this general context
by \citep{ash', kodama} and others. Here, for
comparison, we again assume that the bounding hypersurface $\partial
{\mathcal{M}}$ is diffeomorphic to $S^{3}$. Note further that the
value of $I_{CS}$ for a particular classical solution does not change
as one evolves the boundary data $\omega_{ai}$ in (say) the normal
direction, because of the Hamiltonian (normal) constraint
$\tilde{\sigma}^{i}\tilde{\sigma}^{j}R_{ij}=0$, arising from
Eq.(\ref{asymp}). In the case of Bianchi-IX symmetry, for
Einstein-$\Lambda$ gravity, the corresponding \underline{Chern-Simons
quantum states}
\begin{equation}  
\Psi_{CS}[\omega_{ai}]=\exp(\pm I_{CS}[\omega_{ai}]/\hbar)\label{psics}
\end{equation} 
in quantum cosmology have been further studied
by \citep{luoko, graham, cheng}. For
$N=1$ (simple) supergravity, including a non-zero positive
cosmological constant $\Lambda$ \citep{jacob}, this state has been studied in the case of
$k=+1$ cosmology (round $S^{3}$) by \citep{sano}; see
also \citep{sano'}. In all of these mini-superspace treatments, it is
clear that the Chern-Simons state(s) are at least WKB or
semi-classical approximations to exact quantum states; similarly in
the full theory \citep{ash'}. An excellent review of
Yang-Mills theory in Hamiltonian form, the Yang-Mills Chern-Simons
action and its r\^{o}le in topology and the quantum theory, is given
by Jackiw (1984).

There has been some discussion as to whether the Chern-Simons state
$\Psi_{CS}$ with the minus sign in Eq.(\ref{psics}) is also the
Hartle-Hawking state \citep{luoko, cheng}, at least within the
context of Bianchi-IX symmetry. I doubt whether the last word has been
said on this subject, despite the definite tone of the latter
paper. The argument involves the stability of the SO(4)-spherically
symmetric solution of the evolution equation (\ref{evolution}) for
$\omega_{ai}$; for definiteness, assume here that $\Lambda<0$, giving
a hyperboloid ${\mathbb H}^{4}$ as the maximally symmetric solution. The
corresponding anti-self-dual evolution has the form
$\omega_{ai}=A(t)(\sigma_{a})_{i}$, where $\sigma_{a}$ $(a=1,2,3)$
denotes the orthonormal basis of left-invariant 1-forms on $S^{3}$, as
used in section \ref{example}, and \citep{cheng}
\begin{equation}
A(t)=\frac{1}{2}[1+\cosh(4mt)];\,\,\,\Lambda=-12m^{2}.\label{At}
\end{equation}
Here $t$ is a `Euclidean time coordinate', with its `origin' chosen to
be at $t=0$. Correspondingly, the form of the resulting
$\tilde{\sigma}^{ai}$ shows that the intrinsic radius $a(t)$ of the
$S^{3}$ at `time' $t$ is given by
\begin{equation}
a^{2}(t)=\frac{1}{8m^{2}}\sinh^{2}(4mt),
\end{equation}
as it should be for a 4-space of constant negative curvature. Small
gravitational perturbations of ${\mathbb H}^{4}$, whether of Bianchi-IX type or
generic inhomogeneous distortions, give a linearised unprimed Weyl
spinor $\Psi_{ABCD}$, obeying the linearised Bianchi identity (\ref{weyl3})
in the background ${\mathbb H}^{4}$. The appropriate spherical harmonics
$X_{ABCD}$ and $\bar{Y}_{ABCD}$ on $S^{3}$ are, respectively, of
positive and of negative frequency with respect to the spatial
first-order (Dirac-like) projection of $\nabla_{AA'}$. The linearised
$t$-evolution of such a harmonic is regular at $t=0$ for positive
frequency, but singular for negative frequency. As was seen in section
\ref{example}, there is a one-parameter family of regular anti-self-dual
Einstein metrics, containing the reference ${\mathbb H}^{4}$, of biaxial
Bianchi-IX type. When linearised about ${\mathbb H}^{4}$, they give for
$\omega_{ai}$ or $\Psi_{ABCD}$ a lowest-order (homogeneous)
positive-frequency harmonic, multiplying a function of $t$, with
regular behaviour near the `origin' $t=0$ \citep{luoko}. But generic
linearised data on a bounding $S^{3}$, in the $\omega_{ai}$
description, will give perturbations of $\omega_{ai}$ away from the
background value $A(t)(\sigma_{a})_{i}$, diverging as $t\to 0$ within
a linearised approximation.

Hence, the linear r\'{e}gime does not give enough information, and one
must confront the full non-linear (but first-order) evolution equation
(\ref{evolution}) for $\omega_{ai}$. This partial differential equation is
somewhat reminiscent of the heat-like equation for the Ricci flow on
Riemannian manifolds, studied by Hamilton (1982,1986,1988), and it is
possible that it might be susceptible to related techniques. This is
under investigation; see also \citep{mason}. Of course, there are also
descriptions of the general solution of the Riemannian anti-self-dual
Einstein equations, for $\Lambda\neq 0$, in terms of twistor
theory \citep{ward, wardwells} and in terms of
${\mathcal{H}}$-space \citep{lebrun'}. 

At least, in the much simpler case of (abelian) Maxwell theory, when
one takes the (anti-)self-dual part of the spatial connection (vector
potential) $A_{i}$ to be the `coordinate variables', the normalisable
Chern-Simons state $\Psi_{CS}$ \underline{is} the ground state
\citep{ash''}. This corresponds to the \underline{wormhole state}
in quantum cosmology \\ \citep{hawking88, death}. Similarly, one expects that the non-normalisable
`state' $\Psi_{CS}$, corresponding to the opposite sign in (\ref{psics}),
gives the Maxwell version of the Hartle-Hawking `state'. Note here
that, when the Maxwell field in this representation is split up into
an infinite sum of harmonic oscillators, the description of each
oscillator is that of the \underline{holomorphic representation}
\citep{faddeev}; this recurs in supergravity.

In gravity, the ubiquitous Chern-Simons action $I_{CS}$ of Eq.(\ref{ics'})
re-appears (naturally) as the generating function in the
transformation from `traditional' coordinates $e^{AA'}{}_{i}$ and
conjugate momenta $p_{AA'}{}^{i}$ to Ashtekar variables
$(\omega_{ABi}, \tilde{\sigma}^{ABi})$ \citep{mielke}. The
corresponding property for N=1 (simple) supergravity is described by
Mac\'{i}as (1996).

One might then ask whether, for further generalisations containing Einstein
gravity and other fields, corresponding (Euclidean) actions $I_{CS}$
can be found from descriptions of Ashtekar type. This requires (first)
a suitably `form-al' geometric treatment of the Lagrangian. Robinson (1994, 1995) has done this for, respectively,
Einstein-Maxwell and Einstein-Yang-Mills theory, both with
$\Lambda$-term; see also Gambini and Pullin (1993). For
relations between anti-self-dual Yang-Mills theory and anti-self-dual
gravity, see, for example, \citep{bengtsson}. It would be
extremely interesting if the generality could be increased to include,
for example, N=1 supergravity with gauged supermatter, with gauge
group SU(2), SU(3), $\ldots$ \citep{wessbagger, death}
- -- see the following sections \ref{can} and \ref{AJ}.

\section{Canonical quantum theory of N=1 \\ supergravity: `traditional
variables'}\label{can}

\subsection{Dirac approach}\label{dirac}

\noindent

Turning back to supergravity, consider the Dirac canonical treatment
of simple N=1 supergravity, using the `traditional variables'
$(e^{AA'}{}_{i}, p_{AA'}{}^{i}, \psi^{A}{}_{i},
\tilde{\psi}^{A'}{}_{i})$ \citep[1996]{death84}, which are the
natural generalisation of the `traditional' variables $(e^{AA'}{}_{i},
p_{AA'}{}^{i})$ for Einstein gravity, based on the spatial tetrad components
$e^{a}{}_{i}$ and their conjugate momenta $p_{a}{}^{i}$
$(a=0,1,2,3)$. In the supergravity version, the fermionic quantities
$(\psi^{A}{}_{i}, \tilde{\psi}^{A'}{}_{i})$ are the spatial
projections of the spin-3/2 potentials $(\psi^{A}{}_{\mu},
\tilde{\psi}^{A'}{}_{\mu})$. As classical quantities, they are odd
Grassmann quantities, anti-commuting among themselves, but commuting
with bosonic quantities such as $e^{AA'}{}_{i}$ and $p_{AA'}{}^{i}.$ The
bosonic quantities, such as $e^{AA'}{}_{i}, p_{AA'}{}^{i}$, are not
necessarily Hermitian  complex (in Lorentzian signature, say), but are
generally even elements of a Grassmann algebra; that is, they are
(schematically) of the form (complex)\,+\,(complex)$\times$(bilinear in
$\psi, \tilde{\psi}$)+ analogous fourth-order terms $+\ldots$. The
bosonic fields $e^{AA'}{}_{i}(x), p_{AA'}{}^{i}(x)$ are canonical
conjugates, and the canonical conjugate of $\psi^{A}{}_{i}(x)$, in the
fermionic sense of Casalbuoni (1976), is 
\begin{equation}
\pi_{A}{}^{i}=-\frac{1}{2}\varepsilon^{ijk}\tilde{\psi}^{A'}{}_{j}e_{AA'k}.\label{pi}
\end{equation}

In the quantum theory, one can, for example, consider wave-functionals
of the `traditional' form:
\begin{equation}
\Psi=\Psi[e^{AA'}{}_{i}(x), \psi^{A}{}_{i}(x)],\label{wavefnal}
\end{equation}
living in a Grassmann algebra over the complex numbers ${\mathbb
C}$. Following the Dirac approach, a wave-function $\Psi$, describing
a physical state, must obey the \underline{quantum constraints},
corresponding to the classical constraints appearing in a Hamiltonian
treatment of a theory with local invariances, as seen in section
\ref{ashvar} for Ashtekar variables. Taking the case of Lorentzian
signature, for definiteness, the only relevant quantum constraints to
be satisfied are the \underline{local Lorentz constraints}
\begin{equation}
J^{AB}\Psi=0,\,\,\,\,\,\,\bar{J}^{A'B'}\Psi=0,\label{J}
\end{equation}
together with the \underline{local supersymmetry constraints}
\begin{equation}
S^{A}\Psi =0,\,\,\,\,\,\,\bar{S}^{A'}\Psi =0,\label{S}
\end{equation}
The local Lorentz constraints (\ref{J}) simply require that the
wave-functional $\Psi$ be constructed in a locally Lorentz-invariant
way from its arguments; that is, that all unprimed and all primed
spinor indices be contracted together in pairs. Classically, the
fermionic expression $\tilde{S}_{A'}$ is given by
\begin{equation}
\tilde{S}_{A'}=\varepsilon^{ijk}e_{AA'i}(^{3s}D_{j}\psi^{A}{}_{k})+\frac{1}{2}i\kappa^{2}\psi^{A}{}_{i}p_{AA'}{}^{i},\label{S1}
\end{equation}
with $\kappa^{2}=8\pi$, where $^{3s}D_{j}()$ denotes a suitable
torsion-free spatial covariant derivative \citep[1996]{death84}. The classical $S^{A}$ is given formally by
Hermitian conjugation of (\ref{S1}).

In the quantum theory, the operator $\bar{S}_{A'}$ contains only a
first-order bosonic derivative:
\begin{equation}
\bar{S}_{A'}=\varepsilon^{ijk}e_{AA'i}(^{3s}D_{j}\psi^{A}{}_{k})+\frac{1}{2}\hbar\kappa^{2}\psi^{A}{}_{i}\frac{\delta}{\delta
e^{AA'}{}_{i}}.\label{S2}
\end{equation}
 The resulting constraint,
$\bar{S}_{A'}\Psi =0$, then has a simple interpretation in terms of
the transformation of $\Psi$ under a local primed supersymmetry
transformation, parametrized by $\tilde{\varepsilon}^{A'}(x)$, acting on
its arguments. One finds \citep[1996]{death84} that, under the
supersymmetry transformation with 
\begin{equation}
\delta e^{AA'}{}_{i}=i\kappa\tilde{\varepsilon}^{A'}\psi^{A}{}_{i},
\,\,\,\,\delta\psi^{A}{}_{i}=0, \label{deltae} 
\end{equation}
the change $\delta\Psi$ is given by 
\begin{equation}
\delta(\log\Psi)=\frac{-2i}{\hbar\kappa}\int d^{3}x\,\varepsilon^{ijk}e_{AA'i}(^{3s}D_{j}\psi^{A}{}_{k})\tilde{\varepsilon}^{A'}\label{deltapsi}. 
\end{equation}
The quantum version of $S^{A}$ is more complicated in this
representation, involving a mixed second-order functional derivative,
schematically $\delta^{2}\Psi/\delta e\delta\psi$. However, one can
move between the $(e^{AA'}{}_{i},\psi^{A}{}_{i})$ representation  and
the $(e^{AA'}{}_{i},\tilde{\psi}^{A'}{}_{i})$ representation, by means
of a suitable functional Fourier transform. In the latter
representation, the operator $S^{A}$ appears simple, being of first
order, while the operator $\bar{S}_{A'}$ appears more
complicated. Finally, note that in N=1 supergravity, there is no need
to study separately the quantum constraints ${\mathcal{H}}^{AA'}\Psi
=0$, corresponding to local coordinate invariance in 4 dimensions, and
summarised classically in the Ashtekar representation by the vanishing
of the quantities multiplying the Lagrange multipliers in
Eq.(\ref{S}). This is because the anti-commutator of the fermionic
operators $S^{A}$ and $\bar{S}^{A'}$ gives ${\mathcal{H}}^{AA'}$, in a
suitable operator ordering, plus quantities multiplying $J^{AB}$ or
$\bar{J}^{A'B'}$; hence, the annihilation of $\Psi$ by
$S^{A},\bar{S}^{A'},J^{AB}$ and $\bar{J}^{A'B'}$ implies further that
${\mathcal{H}}^{AA'}\Psi =0$.

\subsection{The quantum amplitude}

Consider, within N=1 simple supergravity, the `Euclidean' quantum
amplitude to go from given asymptotically flat initial data, specified
by $(e^{AA'}{}_{iI}(x),\tilde{\psi}^{A'}{}_{iI}(x))$ on ${\mathbb R}^{3}$, to
given final data $(e^{AA'}{}_{iF}(x),\psi^{A}{}_{iF}(x))$, within a
Euclidean time-separation $\tau >0$, as measured at spatial
infinity. Formally, this is given by the path integral
\begin{equation}
K(e_{F},\psi_{F};e_{I},\tilde{\psi}_{I};\tau)=\int\exp(-I/\hbar){\mathcal{D}}e{\mathcal{D}}\psi
{\mathcal{D}}\tilde{\psi}, \label{K}
\end{equation}
where $I$ denotes a version of the Euclidean action of supergravity,
appropriate to the boundary data \citep[1996]{death84}, and Berezin
integration is being used for the fermionic variables \\ \citep{faddeev}. Of course, this is very close to being
a Hartle-Hawking integral, as in (\ref{psihh}), except that part of the
boundary has been pushed to spatial infinity, and that the fermionic
data have been taken in different forms on the initial and final
${\mathbb R}^{3}$ .

As in any theory with local (gauge-like) invariances, when treated by
the Dirac approach, the quantum constraint operators at the initial and final
surfaces annihilate the quantum amplitude $K$ above. In particular, on
applying (say) the supersymmetry constraint \\ $\bar{S}_{A'}K=0$ at the
final surface, one obtains
\begin{equation}
\varepsilon^{ijk}e_{AA'iF}(^{3s}D_{j}\psi^{A}{}_{kF})K+\frac{1}{2}\hbar\kappa^{2}\psi^{A}{}_{iF}\frac{\delta
K}{\delta e^{AA'}{}_{iF}}=0.\label{K1}
\end{equation}
As in section \ref{dirac}, this describes how $K$ changes (in a simple
way) when a local primed supersymmetry transformation (\ref{deltae}) is
applied to the final data $(e^{AA'}{}_{iF},\psi^{A}{}_{iF}).$ One then
considers the semi-classical expansion of this `Euclidean' quantum
amplitude, by analogy with Eq.(\ref{asymp}). However, one should first note
that, in general, there is no classical solution
$(e^{AA'}{}_{\mu},\psi^{A}{}_{\mu},\tilde{\psi}^{A'}{}_{\mu})$ of the
supergravity field equations, agreeing with the initial and final data
as specified above, and corresponding to a Euclidean time interval
$\tau$ at spatial infinity. This was not appreciated in
\citep{death84}, but was later corrected in \citep{death}. The
difficulty resides in the classical $\tilde{S}^{A'}=0$ constraint at
the final surface [Eq.(\ref{S1})], and similarly $S^{A}=0$ at the
initial surface; it is precisely related to the primed supersymmetry
behaviour of Eqs.(\ref{deltae},8) finally, and similarly for unprimed
supersymmetry initially.

Suppose now that we start from a purely bosonic (Riemannian) solution
$e^{AA'}{}_{\mu}$ of the vacuum Einstein field equations, while
$\psi^{A}{}_{\mu}=0$, $\tilde{\psi}^{A'}{}_{\mu}=0$. Then, for the
corresponding bosonic boundary data $e^{AA'}{}_{iI}$ and
$e^{AA'}{}_{iF}$, we expect there to exist a semi-classical expansion
of $K(e_{F},0;e_{I},0;\tau)$, as in Eq.(\ref{asymp}). By studying (say) the
quantum supersymmetry constraint  $\bar{S}_{A'}K=0$ of Eq.(\ref{K1}) at
the final surface, and allowing the variable $\psi^{A}{}_{iF}(x)$ at
the final surface to become small and non-zero, while still obeying the
\underline{classical} $\tilde{S}_{A'}=0$ constraint (\ref{S1}) at the
final surface, one finds:
\begin{equation}
A_{0}=const., \,\,\,\,A_{1}=A_{2}=\ldots =0,\label{A0}
\end{equation}
for the loop prefactors $A_{0}, A_{1}, A_{2},\ldots$ in the expansion
(\ref{asymp}). Thus, in N=1 supergravity, for purely bosonic boundary data,
the semi-classical expansion of the `Euclidean' amplitude $K$ is
\underline{exactly semi-} \underline{classical}; that is,
\begin{equation}
K\sim A_{0}\exp(-I_{class}/\hbar),\label{Kbos}
\end{equation}
where $A_{0}$ is a constant. The symbol $\sim$ for an asymptotic
expansion, has been used in Eq.(\ref{Kbos}), rather than equality $=$, as
there will sometimes be more than one inequivalent \underline{complex}
solution of the vacuum Einstein equations joining $e^{AA'}{}_{iI}$ to
$e^{AA'}{}_{iF}$. In that case, there will be more than one classical
action $I_{class}$, but only the leading contribution, corresponding
to the most negative value of Re$[I_{class}]$, will appear in
Eq.(\ref{Kbos}). Since there has been some disagreement in the past about
the result described in this paragraph, it should be noted that no published
paper, since the publication of the revised argument in
\citep{death}, has given a substantive contrary argument.

Finally, one might ask whether more general amplitudes \\
$K(e_{F},\psi_{F};e_{I},\tilde{\psi}_{I};\tau)$ in N=1 supergravity share some
of the simplicity of the purely bosonic amplitude above. Here,
non-trivial fermionic data $\psi^{A}{}_{F}$ and $\tilde{\psi}^{A'}{}_{I}$
should be chosen such that there is a classical solution joining the
data in Euclidean time $\tau$, whence a semi-classical expansion of
$K$ should exist, by analogy with Eq.(\ref{asymp}). By considering the
possible form of locally supersymmetric counterterms, formed from
volume and surface integrals of various curvature invariants, at
different loop orders, one is led to expect that the full amplitude
$K$ might well be finite on-shell; certainly, the purely bosonic part
of each invariant must be identically zero, by the property (\ref{A0}),
and it would then be odd if some of its partners, which are quadratic,
quartic, $\ldots$ in fermions, managed not to be zero identically. A
more detailed investigation of fermionic amplitudes, at one loop in
N=1 supergravity with gauged supermatter, is given in
\citep{death99}.

\subsection{N=1 supergravity with gauged supermatter}\label{N}

\indent

Gauge theories of `ordinary matter', with spins $0, \frac{1}{2},1,$
can be combined with N=1 supergravity (say) in  a very geometrical
way, to give a theory with \underline{four types of local
invariance}:-- local coordinate, local tetrad rotation, local N=1
supersymmetry, and (say) local SU$(n)$ invariance
\citep{wessbagger}. The resulting theory is uniquely defined once
the coupling constant $g$ is specified (with $g^{2}=1/137.03\ldots$),
except for an analytic potential $P(a^{I})$, where the $a^{I}$ are
complex scalar fields, which live on complex projective space
${\mathbb C}P^{n-1}$ (for $n\geq 2$). In the simplest non-trivial
case, with SU(2) gauge group, there is one complex scalar field $a$,
with complex conjugate $\bar{a}$. There is a natural
\underline{K\"{a}hler metric} on ${\mathbb C}P^{1}$, or the Riemann
sphere, parametrised by $a$ (provided one includes the point
$a=\infty$ at the North pole, while $a=0$ corresponds to the South
pole). The \underline{K\"{a}hler potential} is
\begin{equation}
K=\log(1+a\bar{a}),\label{kahler}
\end{equation} 
giving the \underline{K\"{a}hler metric}
\begin{equation}
g_{11^{*}}=\frac{\partial^{2}K}{\partial a\partial\bar{a}}=\frac{1}{(1+a\bar{a})^{2}}.\label{kahler1}
\end{equation} 
Equivalently, this metric reads as 
\begin{equation}
ds^{2}=\frac{da\,d\bar{a}}{(1+a\bar{a})^{2}},\label{kahler2}
\end{equation} 
which is the metric on the unit round 2-sphere (really, ${\mathbb
C}P^{1}$). Not surprisingly, the isometry group for this geometry is
just the original gauge group SU(2).

The other fields in the  SU(2) theory may be summarised as
follows. There is a spin-1/2 field $(\chi_{A},\tilde{\chi}_{A'})$,
which has no Yang-Mills index in this case, and which is the partner
of $(a,\bar{a})$. The Yang-Mills potential(s) $v^{(a)}_{\mu}$, with
$(a)=1,2,3,$ have fermionic  spin-1/2 partners
$(\lambda^{(a)}_{A},\tilde{\lambda}^{(a)}_{A'})$; thus there is a
distinction between two different types of underlying  spin-1/2 field
- -- the $\chi$'s and the $\lambda$'s. As usual, gravity is described by
the tetrad $e^{AA'}{}_{\mu}$, with spin-3/2 supersymmetry partner
$(\psi^{A}_{\mu},\tilde{\psi}^{A'}_{\mu})$. The relevant Lagrangian
may be found in \citep{wessbagger}.

This model can be extended to the group SU(3), for example, by using
the corresponding K\"{a}hler metric given in \citep{gibbpope}. A
suitable basis of 8 generators of the Lie algebra $\mathfrak
s\mathfrak u$(3), as employed in the Lagrangian of
\citep{wessbagger}, is given by the Gell-Mann matrices in \citep{itzykson}.

Perhaps the most immediately striking feature of the resulting
Lagrangian is the enormous \underline{negative cosmological constant}
$\Lambda$, with
\begin{equation}
\Lambda =-\frac{g^{2}}{8}\label{lambda}
\end{equation}  
in Planck units, for SU(2) and SU(3). However, at least in the case of
zero potential [$P(a)=0$], when the theory is written out in
Hamiltonian form, and the Dirac approach to
canonical quantisation is taken \citep[1999]{death}, then the N=1
local supersymmetry again implies that quantum amplitudes $K$ to go
between initial and final \underline{purely bosonic configurations} in
a Euclidean time-at-infinity $\tau$ are \underline{exactly
semi-classical:}
\begin{equation}
K\sim const.\exp(-I_{class}/\hbar).\label{Kbos1}
\end{equation}
Correspondingly, one might again expect some related simplification in
amplitudes $K$ for which there are non-trivial \underline{fermionic}
boundary data, in addition to gravity, Yang-Mills and scalars. Some
evidence of this has been found in an investigation of one-loop
corrections in the SU(2) theory, where the `background' purely bosonic
classical solution is taken to be a suitable hyperboloid ${\mathbb H}^{4}$, corresponding to the negative value of $\Lambda$, and the
`unperturbed' initial and final boundaries are taken to be two round
3-spheres $S^{3}$ at different radii from the `centre' of the ${\mathbb
H}^{4}$ \citep{death99}. Consistent weak-field fermionic boundary
data are put on the spheres, and the one-loop corrections to the
quantum amplitude are studied with the help of \underline{both} (local)
quantum supersymmetry constraint operators $S^{A}$ and $\bar{S}^{A'}.$
The Dirac approach to the computation of loop terms in such a locally
supersymmetric theory was found to be extremely streamlined by
comparison with the corresponding path-integral
calculation. Typically, fermionic one-loop examples of this type are
often very simple, sometimes not even involving an infinite sum or
integral. In non-trivial examples, the amplitudes appear to be
exponentially convergent \citep{death99}, and the structure suggests
that this will continue at higher loop order.

Since the loop behaviour of this SU(2) model appears reasonable, it
would seem worthwhile to investigate this and other SU$(n)$ models
further, to understand better their physical consequences, and to try
to predict effects which are observable at accelerator energies.

\section{Canonical quantisation of N=1 supergravity: Ashtekar-Jacobson
variables}\label{AJ}

\noindent

Now consider the canonical variables introduced by Jacobson (1988) for
N=1 supergravity, following Ashtekar's approach, possibly including a
\underline{positive cosmological constant} written as $\Lambda
=12\mu^{2}$ [cf. Eq.(\ref{At}) for negative $\Lambda$]. For consistency,
the fermionic variables have been re-normalised as in D'Eath
(1996). The bosonic variables are again taken to be the connection
1-forms $\omega_{ABi}=\omega_{(AB)i}$, together with the
canonically-conjugate variables
$\tilde{\sigma}^{ABi}=\tilde{\sigma}^{(AB)i}$. The fermionic variables
are taken to be the unprimed spatial 1-forms $\psi_{Ai}$ and their
conjugate momenta $\tilde{\pi}^{Ai}$. Once again, all variables only
involve unprimed spinor indices, and so are well adapted to a
treatment of (anti-)self-duality. Note that $\tilde{\pi}^{Ai}$ is
given in terms of the `traditional' variables of section \ref{can} by 
\begin{equation}
\tilde{\pi}^{Ai}=\frac{i}{\sqrt{2}}\varepsilon^{ijk}e^{AA'}{}_{j}\tilde{\psi}_{A'k}.\label{pi6}
\end{equation}

The classical supersymmetry constraints involve
\begin{eqnarray}
S^{A} & = & {\mathcal{D}}_{i}\tilde{\pi}^{Ai}+4i\mu(\tilde{\sigma}^{k}\psi_{k})^{A}\nonumber\\
& = & 0,\label{SA}
\end{eqnarray}
where ${\mathcal{D}}_{i}$ is a spatial covariant derivative involving
the connection $\omega_{ABi}$, and 
\begin{eqnarray}
S^{\dagger A} & = & (\tilde{\sigma}^{j}\tilde{\sigma}^{k}{\mathcal{D}}_{[j}\psi_{k]})^{A}-4i\mu(\tilde{\sigma}_{k}\tilde{\pi}^{k})^{A}\nonumber\\
& = & 0.\label{SA+}
\end{eqnarray}
Note here that
$\tilde{\sigma}^{AB}{}_{k}=(1/\sqrt{2})\varepsilon_{kmn}\tilde{\sigma}^{ACm}\tilde{\sigma}_{C}{}^{Bn}$
\\ \cite[1988]{jacob, death}.
Quantum-mechanically, for a wave-functional $\Psi[\omega_{ABi}, \psi_{Ai}]$, the constraint $S^{A}\Psi=0$ is a first-order functional differential equation, namely:
\begin{equation}
{\mathcal{D}}_{i}\left(\frac{\delta\Psi}{\delta\psi_{Ai}}\right)-4\mu\psi^{B}{}_{k}\left(\frac{\delta\Psi}{\delta\omega_{ABk}}\right)=0.\label{fde}
\end{equation}
This simply describes the invariance of the wave-functional $\Psi$
under a local unprimed supersymmetry transformation, parametrised by
$\varepsilon^{A}(x)$, applied to its arguments $\omega_{ABi}(x),
\psi_{Ai}(x)$. Note that the unprimed transformation properties of
`traditional' variables include
\begin{equation}
\delta e_{AA'i}=-i\tilde{\psi}_{A'i}\varepsilon_{A},\,\,\,\delta\psi_{Ai}=2{\mathcal{D}}_{i}\varepsilon_{A},\,\,\,\delta\tilde{\psi}_{A'i}=0.\label{trans}
\end{equation}
One further deduces, following \citep{capovilla'}, the variation
\begin{equation}
\delta\omega_{ABi}=\mu\psi_{(Ai}\varepsilon_{B)}.\label{deltaomega}
\end{equation}

However, the quantum constraint $S^{\dagger A}\Psi=0$ is described by
a complicated second-order functional differential equation. One can
(say) transform from `coordinate' variables $(\omega_{ABi},\psi_{Ai})$
to the opposite `primed' coordinates \\
$(\tilde{\omega}_{A'B'i},\tilde{\psi}_{A'i})$, {\it via} `traditional'
coordinates $(e_{AA'i},\psi_{Ai})$ \citep{macias}, using
functional Fourier transforms \citep{death}, with Berezin
integration over fermionic variables \citep{faddeev}. In the
`primed' coordinates $(\tilde{\omega}_{A'B'i}, \tilde{\psi}_{A'i})$,
the quantum constraint operator $S^{A}$ will appear complicated and
second-order, while the operator $S^{\dagger A}$ becomes simple and
first-order.

In the unprimed representation $(\omega_{ABi},\psi_{Ai})$, in the case
$\Lambda =12\mu^{2}>0$, one can again define the Chern-Simons action
$I_{CS}$ for N=1 supergravity \citep{sano, sano'} as:
\begin{equation}
I_{CS}[\omega_{ABi},\psi_{Ai}]=\frac{3}{2\Lambda}\int\,W,\label{ics}
\end{equation}
\begin{equation}
W=\omega_{AB}\wedge d\omega^{AB}+\frac{2}{3}\omega_{AC}\wedge\omega^{C}{}_{B}\wedge\omega^{AB}-\mu\psi^{A}\wedge
{\mathcal{D}}\psi_{A}.\label{W}
\end{equation}
Here, we assume that the integration is over a compact (boundary)
3-surface. The notation ${\mathcal{D}}\psi_{A}$ denotes the covariant
exterior derivative of $\psi_{Ai}$, using the connection
$\omega_{ABi}$ \citep{death}. Note that the functional
$I_{CS}[\omega_{ABi},\psi_{Ai}]$ is \underline{invariant under
unprimed local supersy-} \\ \underline{mmetry transformations} (\ref{trans}), (\ref{deltaomega})
applied to its arguments, with parameter $\varepsilon^{A}(x)$.

Correspondingly, the \underline{Chern-Simons wave function},
\begin{equation}  
\Psi_{CS}=\exp(- I_{CS}[\omega_{ABi},\psi_{Ai}]/\hbar),\label{psics'}
\end{equation}
obeys the first quantum supersymmetry constraint
\begin{equation}  
S^{A}\Psi_{CS}=0\label{psics''}.
\end{equation}
But, by symmetry, when one transforms this wave function into the
opposite primed $(\tilde{\omega}_{A'B'i},\tilde{\psi}_{A'i})$
representation, it will have the same form, and hence is also
annihilated by the $S^{\dagger A}$ constraint operator. Hence, since
$\Psi_{CS}$ is automatically invariant under local tetrad rotations,
this Chern-Simons wave function obeys all the quantum constraints, and
so defines a physical `state'. Here inverted commas have been used,
since it is not clear whether or not $\Psi_{CS}$ is normalisable.

The classical action $I_{CS}[\omega_{ABi},\psi_{Ai}]$ is the
generating function for anti-self-dual evolution of boundary data
$(\omega_{ABi}(x),\psi_{Ai}(x))$ given on the compact spatial
boundary, just as $I_{CS}[\omega_{ABi}]$ generated the classical
evolution for Einstein gravity with a non-zero $\Lambda$ term in
section \ref{chern}. This certainly justifies further investigation.

\section{Comments}

\noindent

Dirac's approach to the quantisation of constrained Hamiltonian
systems can be applied to boundary-value problems, whether the
Hartle-Hawking path integral of quantum cosmology, or the transition
amplitude to go from given initial to final asymptotically-flat data
in Euclidean time $\tau$. When the crucial ingredient of local
supersymmetry is added, the main quantum constraints to be satisfied
by the wave-functional $\Psi$ become the supersymmetry constraints
$S^{A}\Psi =0$ and $\bar{S}^{A'}\Psi =0$, each of which is only of
first order in bosonic derivatives. Using `traditional' canonical
variables, one finds that, both for N=1 simple supergravity and for
N=1 supergravity with gauged SU($n$) supermatter (at least in the
simplest case of zero potential), transition amplitudes to go from
initial to final purely bosonic data are exactly semi-classical,
without any loop corrections. This, in turn, strongly suggests that
quantum amplitudes including fermionic boundary data are also finite
in these theories.

Ashtekar's different (essentially spinorial) choice of canonical
variables, for Einstein gravity with cosmological constant $\Lambda$,
allows a very efficient treatment of (anti-)self-dual Riemannian
`space-times'. For $\Lambda\neq 0$, anti-self-dual evolution arises
from the Chern-Simons generating functional $I_{CS}$ of section \ref{chern}. Jacobson's extension of Ashtekar canonical variables to
include N=1 supergravity for $\Lambda\geq 0$ is again adapted to the
study of anti-self-dual supergravity in the Riemannian case. For
$\Lambda> 0$, anti-self-dual evolution in N=1 supergravity similarly
arises from the Chern-Simons functional $I_{CS}$ of section \ref{AJ}. Further, when Ashtekar-Jacobson variables are used, the
Chern-Simons wave function $\Psi_{CS}=\exp(-I_{CS})$, of N=1
supergravity with $\Lambda> 0$, gives an exact solution of all the
quantum constraints. (In contrast, it is not clear whether such a
statement is meaningful in the non-supersymmetric case of section
 \ref{chern}.) There remains the difficult question of the relation
between $\Psi_{CS}$ and the Hartle-Hawking state.

Such inter-connections between (anti-)self-duality in Riemannian
geometry and meaningful states in quantum cosmology may well provide
an enduring Union (or, possibly, an Intersection) between Oxford and
Cambridge and other such Centres of Gravity.

\section*{Acknowledgements}

I am very grateful to Andrew Farley, Gary Gibbons, Ted Jacobson, John
Thompson and Pelham Wilson for valuable discussions.  

\newpage

\section*{References}
\bibliographystyle{alpha}
%\bibliography{deathbib}

\end{document}